\documentclass[prl,twocolumn]{revtex4}

\usepackage{graphicx}
\usepackage{dcolumn}
\usepackage{bm}

\DeclareGraphicsExtensions{.eps,.gif,.ps}

\begin{document}


\title{Relativistic effects in quantum walks: Klein's paradox and Zitterbewegung}

\author{Pawe{\l} Kurzy\'nski}

\email{kurzpaw@hoth.amu.edu.pl}

\affiliation{Faculty of Physics,
 Adam Mickiewicz University,
Umultowska 85, 61-614 Pozna\'{n}, Poland.}

\date{June 20, 2006}%

\begin{abstract}
Quantum walks are not only algorithmic tools for quantum computation
but also not trivial models which describe various physical
processes. The paper compares one-dimensional version of the free
particle Dirac equation with discrete time quantum walk (DTQW). We
show that the discretized Dirac equation when compared with DTQW
results in interesting relations. It is also shown that two
relativistic effects associated with the Dirac equation, namely
Zitterbewegung (quivering motion) and Klein's paradox, are present
in DTQW, which can be implemented within non-relativistic quantum
mechanics.
\end{abstract}

\maketitle

\newcommand{\ga}{\gamma}
\newcommand{\la}{\lambda}
\newcommand{\ran}{\rangle}
\newcommand{\lan}{\langle}

Since the papers of Aharonov {\it{et al.}} \cite{Ahar} and Farhi and
Gutmann \cite{Farhi}, quantum walks have been deeply investigated in
hope to find faster algorithms (Refs.~\cite{Kempe, Ambainis} and
references therein). Despite their contribution to the quantum
information processing, quantum walks are themselves very
interesting physical systems worth being studied due to effects from
various fields of physics like quantum chaos \cite{Wojc} and solid
state physics \cite{My}. Recently, it has been shown that discrete
time quantum walk (DTQW) resemble the one-dimensional free particle
Dirac equation \cite{Str1, Bracken}. Actually, one has to keep in
mind that the idea of quantum walks goes back to Feynmann and Hibbs
\cite{Feynmann} who considered a discrete version of the
one-dimensional Dirac equation propagator. In this paper we study
differences and similarities of the two models and show that two
effects associated with the Dirac equation, Klein's paradox and
Zitterbewegung, occur in DTQW. This is an important result leading
to the fact that the Dirac equation might be simulated via DTQW.

The one-dimensional free particle Dirac equation has the form
\begin{equation} \label{e1}
i \frac{\partial}{\partial t} \psi(x,t) = -i \sigma_i
\frac{\partial}{\partial x} \psi(x,t) + m \sigma_j \psi(x,t),
\end{equation}
where $m$ is a mass of the particle, and $\sigma_i$ and $\sigma_j$
have to obey $\sigma_i^2=\sigma_j^2=I$ and $\sigma_i \sigma_j +
\sigma_j \sigma_i=0$, as we want Eq.~(\ref{e1}) to be Lorentz
invariant. We assume $c=\hbar=1$ here and throughout the paper. The
most natural way is to take $\sigma_i$ and $\sigma_j$ to be the two
distinct Pauli matrices. In this case the wave function has two
components $\psi(x,t)=\left( \psi_u(x,t), \psi_d(x,t) \right)^T$.
The eigenvalues of Eq.~(\ref{e1}) are $E_{\pm}=\pm \sqrt{k^2+m^2}$,
where $k$ is the eigenvalue of the momentum operator. The striking
feature of the Dirac equation is that it gives positive and negative
energy solutions and that there is a gap of forbidden energy region
$-m<E<m$.

Now, let us define DTQW. The step operator of one-dimensional DTQW
is given by $U=TC$, where
\begin{equation} \label{e2}
T=e^{ip_x \sigma_z l}
\end{equation}
is the conditional shift operator indroduced by Aharonov {\it{et
al.}} \cite{Ahar}, and $C$ is a two-dimensional coin operator we
assume in the form
\begin{equation} \label{e3}
C= e^{i \frac{\theta}{2}\vec{n} \cdot \vec{\sigma} }=\left(
\begin{array}l \alpha_+~~\beta_+ \\ \alpha_-~~\beta_-
\end{array} \right).
\end{equation}
In Eq.~(\ref{e2}), $p_x$, $\sigma_z$ and $l$ are the $x$ component
of momentum operator, the $z$ component of spin vector operator and
the size of one step respectively. In Eq.~(\ref{e3}), $\theta$ is
the angle of rotation about $\vec{n}=(n_x,n_y,n_z)^T$, which is an
arbitrary unit vector, and
$\vec{\sigma}=(\sigma_x,\sigma_y,\sigma_z)^T$ is the spin vector
operator consisting of three Pauli matrices, therefore
$\alpha_{\pm}=\cos{\frac{\theta}{2}} \pm i n_z
\sin{\frac{\theta}{2}}$ and $\beta_{\pm}=(in_x \pm
n_y)\sin{\frac{\theta}{2}}$. The state of the walk is described by
the two component wave function
$\psi(x,t)=(\psi_+(x,t),\psi_-(x,t))^T$, like in the one-dimensional
Dirac equation, and from Eqs.~(\ref{e2}) and (\ref{e3}) one can
obtain a recursive formula
\begin{equation} \label{e4}
\psi_{\pm}(x,t+\Delta t)=\alpha_{\pm}\psi_{\pm}(x\mp l,t) +
\beta_{\pm}\psi_{\mp}(x\mp l,t),
\end{equation}
where $\Delta t$ is the time of one step. Taking
$\vec{n}=(\frac{1}{\sqrt{2}},\frac{1}{\sqrt{2}},0)^T$ and
$\theta=\pi$ one gets the well studied Hadamard walk \cite{Kempe}.

\begin{figure}
\scalebox{1.0} {\includegraphics[width=8truecm]{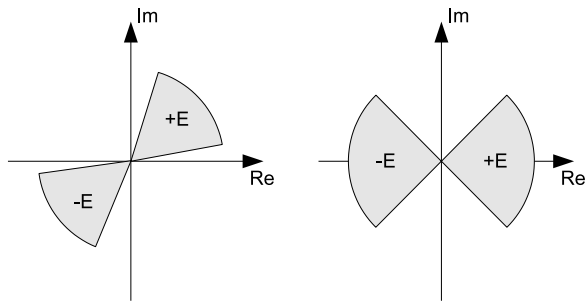}}
\caption{\label{f1} Eigenvalues of DTQW evolution operator on the
complex plane. $+E$ and $-E$ denotes the regions of positive and
negative energy respectively.}
\end{figure}

To calculate eigenvalues of $U$ we have to realize that it commutes
with the translation operator and therefore with the momentum
operator. Thus, the eigenvectors are of the form
$|\lambda\rangle=N(\sum_x e^{ikx} |x\rangle) \otimes |c\rangle$,
where $|c\rangle$ is a position independent state of the coin and
$N$ is a normalization constant. Together with the coin operator
$C$, Eq.~(\ref{e3}), this gives eigenvalues of the form
\begin{eqnarray} \label{e5}
\lambda_{\pm}(k)=\cos{\frac{\theta}{2}}\cos{k}+n_z
\sin{\frac{\theta}{2}}\sin{k} \pm \nonumber \\
i\sqrt{1-(\cos{\frac{\theta}{2}}\cos{k}+n_z
\sin{\frac{\theta}{2}}\sin{k})^2}.
\end{eqnarray}
It is easily verified that $\lambda_{\pm}^{\ast}\lambda_{\pm}=1$ and
one can write them as
$\lambda_{\pm}(k)=e^{-i\varphi_{\pm}(k)}=e^{-iE_{\pm}(k)\Delta t}$,
therefore we may assume that $\lambda_{+}(k)$ and $\lambda_{-}(k)$
correspond to positive and negative energies respectively. On the
complex plane, $\lambda_{\pm}(k)$ are confined to two symmetrically
inverted cones, one corresponding to positive while another to
negative energies (see Fig.\ref{f1}). The dilation angle of the
cones and the angle between the real axis and the line passing
through the middle of the both cones depend on $\theta$ and $n_z$.
For the Hadamard walk, the both angles are $\pi/2$. The region of
complex unit circle not covered by $\lambda_{\pm}(k)$ might be
thought of as a forbidden region. One sees that the above properties
make DTQW and the Dirac Equation alike.

The continuous limit of DTQW, both in time and space, has been
studied in Refs.~\cite{Asl, Bracken, Str1}. It was shown that the
continuous version of DTQW gives the evolution similar to the one
described by the one-dimensional Dirac equation and that both models
give the typical two horn probability distribution for initially
highly localized wave packets. Aslangul \cite{Asl} has studied the
model with a coin of an arbitrary dimension with the coin operator
being rotation of spin about $y$ axis, so the exact comparison with
the Dirac equation cannot be made, although the models seems alike.
It is also important to notice that the coin operator makes DTQW to
be more general than the Dirac equation. In our case it can be an
arbitrary $2 \times 2$ unitary matrix. Due to this fact, the
continuous version of DTQW must not always be Lorentz invariant.
This is the case of the Hadamard walk.

Now, we will do something opposite to what was done in
Refs.~\cite{Asl, Bracken, Str1}. Let us discretize Eq.~(\ref{e1})
and see if it gives Eq.~(\ref{e4}). We take $\sigma_i$ and
$\sigma_j$ to be the $z$ and $y$ Pauli matrices respectively and use
the same time symmetric procedure as in Ref.~\cite{Wess}. We obtain
a recursive formula that conserves probability to the second order
in $\Delta t$:
\begin{equation} \label{e6}
\psi_{{~}_{d}^{u}}(x,t+\Delta t)=\frac{1-\gamma^2}{1+\gamma^2}
\psi_{{~}_{d}^{u}}(x\mp \Delta x,t) \mp \frac{2\gamma}{1+\gamma^2}
\psi_{{~}_{u}^{d}}(x,t),
\end{equation}
where $\gamma=m\Delta t/2$. This looks similar to Eq.~(\ref{e4}),
but to make it more alike one should shift the right hand side of
Eq.~(\ref{e6}) by $l=\Delta x/2$ with the conditional translation
operator $T$, Eq.~(\ref{e2}). It means that DTQW is the discretized
Dirac evolution viewed in, somehow strange, conditional moving
reference frame. Note, that this shift does not occur when one
considers the continuous limit of DTQW. This is due to the fact that
in the continuous limit the differences between $U=TC$ and $U=CT$
vanishes because $C$ and $T$ act simultaneously. In the discrete
regime it is important wether we act with $TC$ or $CT$. One can also
easily check, that in the case $U=CT$, the left hand side of
Eq.~(\ref{e6}) should be shifted with the operator $T^{\dagger}$ in
order to resemble the DTQW recursive formula. Also, choosing
$\sigma_i$ and $\sigma_j$ to be different than here, gives the same
effect. One also needs to remember that since DTQW is by definition
discrete in space and time, it should be rather compared with the
discrete Dirac equation than with its continuous more common
version. It is due to the effects like Bloch oscillations \cite{My}
caused by discrete space and perhaps due to the other effects, yet
unknown, related to discrete time.

\begin{figure}
\scalebox{1.0} {\includegraphics[width=8truecm]{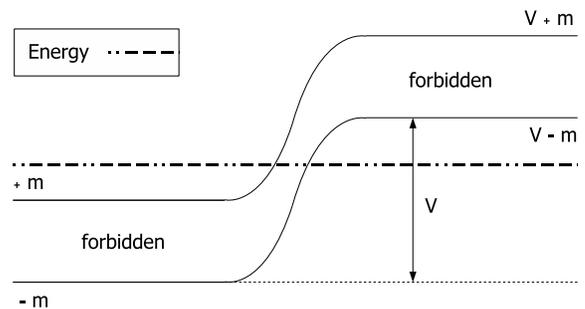}}
\caption{\label{f2} Klein's paradox. The forbidden region $-m<E<m$
is shifted by a potential $V$. The particle with positive energy E
is transferred to the negative energy region. The wave function of
particle is not damped in the new region.}
\end{figure}

Klein's paradox and Zitterbewegung are effects that were first
observed for the Dirac equation (for reference, look for example
Ref.~\cite{Sakurai}). The first one is transfer of the particle with
energy $E$ from the region with zero potential to the region with
potential $V>V-m>E$ (see Fig.\ref{f2}).
In relativistic case the wave function is not damped inside the
potential region, unlike the solutions of the Schr\"odinger
equation. The second effect is the rapid oscillation of a wave
packet due to interference of its positive and negative energy
components.

To observe Klein's paradox in DTQW, one has to introduce a potential
$V$. For simplicity, we take the step potential: all over the region
$x>a$ the potential is $V_0$ and in the region $x \leq a$ the
potential is zero. Adding the uniform potential simply shifts the
energy $E'=E+V_0$ and in the exponential form it may be written as
$e^{-iE'\Delta t}=e^{-i(E+V_0)\Delta t}=e^{-iE\Delta t}e^{i
\varphi}$. From now on, we will identify the potential with
$e^{i\varphi}$. This will be very helpful to describe the existence
of Klein's paradox in DTQW in geometric way.

\begin{figure}
\scalebox{1.0} {\includegraphics[width=8truecm]{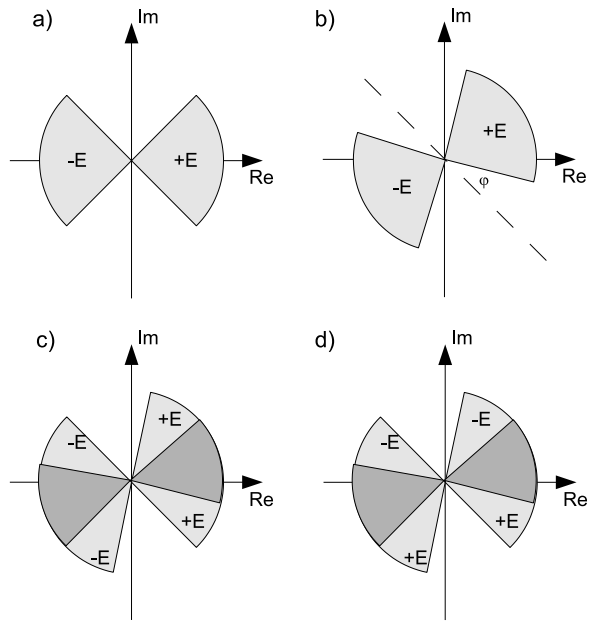}}
\caption{\label{f3} Eigenvalues of DTQW evolution operator on the
complex plane: a) eigenvalues without potential, b) the same
eigenvalues in the presence of a potential $e^{i\varphi}$, c)
intersection of the eigenvalues of two regions, one without
potential and one with the potential $e^{i\varphi}$, giving non
decaying wave functions (dark grey), d) intersection of the
eigenvalues of two regions, one without potential and one with the
potential $e^{i(\varphi+\pi)}$, giving non decaying wave functions
and leading to Klein's paradox (dark grey).}
\end{figure}

The eigenvalues without the potential are presented in Fig.\ref{f3}
a. In the presence of the potential, the eigenvalues are rotated by
the angle $\varphi$, see Fig.\ref{f3} b.
Imagine that DTQW starts in the region without the potential and
that there is a wave packet heading for the potential step
$e^{i\varphi}$. The transmitted part of the wave packet may behave
in two distinct ways: either move further without damping, or start
to decay exponentially fast. Whether it is the first or the second
case, depends directly on the eigenfunction decomposition of the
wave packet. Let $\{ \lambda \}_{0}$ denote the set of the
eigenvalues of DTQW with no potential and $\{ \lambda \}_{\varphi}$
be the set of the eigenvalues in the presence of the potential.
Moreover, let $\{\lambda\}_{wp} \subset \{ \lambda \}_{0}$ be the
set of the eigenvalues which correspond to the eigenfunctions that
contribute to the initial wave packet. By calculating
$\{\lambda\}_{wp} \cap \{ \lambda \}_{\varphi}$ one obtains the part
of the wave packet that will not undergo damping. Now, let us study
how this depends on $\varphi$. For small $\varphi$ the wave packet
is not damped if its energy is higher than the potential step,
similarly as in non-relativistic quantum mechanics, see Fig.\ref{f3}
c. As $\varphi$ rises, more eigenvalues fall into the forbidden
region and the corresponding eigenfunctions are damped, until
$\varphi$ is large enough and the eigenvalues from the cone
$E_{+}$($E_{-}$) intersect with the eigenvalues of the rotated cone
$E_{-}$($E_{+}$) (Fig.\ref{f3} d). This is exactly Klein's paradox.
We should also point out that for specific coins, namely for the
coins which make the eigenvalues cone dilation angle less or equal
to $\pi/2$, one can choose a potentials that damp all
eigenfunctions. The Hadamard walk is always damped in the potential
region if $\varphi=\pi/2$. This might be useful if one would like to
confine DTQW to some region. For example, it is possible to study
the quantum walk on the line with one or two reflecting walls.

Zitterbewegung oscillation are expected to occur in both position
and velocity. To measure this effect, we calculate how the position
changes with the one step $\Delta X=X(t+\Delta t)-X(t)$. This is the
discrete version of velocity. In the Heisenberg picture this might
be written as $\Delta X=U^{\dagger}X U - X$, where $X=\sum_x x
|x\rangle \langle x|$ and $U=TC$, as before. This leads us to the
position independent coin operator
\begin{equation} \label{e7}
\Delta X = \left( \begin{array}l ~|\alpha_+|^2-|\beta_-|^2
~~~~~~\alpha_+^{\ast} \beta_+ - \alpha_- \beta_-^{\ast}\\
\alpha_+ \beta_+^{\ast} - \alpha_-^{\ast}
\beta_-~~~~~~|\beta_+|^2-|\alpha_-|^2
\end{array} \right).
\end{equation}
To calculate its time dependence, we perform the discrete Fourier
transform on the initial state and use the momentum representation
to derive the expectation value of $\Delta X$ as a function of time.
Calculation of DTQW probability distribution via Fourier transform
was presented by Nayak and Vishwanath \cite{Nayak}. The general form
of the time dependent wave function in momentum representation is
\begin{equation} \label{e8}
\tilde{\psi}(k,t) = \sum_{j=\pm}
f_{j}(k)(\lambda_{j}(k))^t|c_{j}(k)\rangle,
\end{equation}
where $f_{\pm}(k)$ are any square integrable functions obeying
$\int_{-\pi}^{+\pi} dk(|f_{+}(k)|^2+|f_{-}(k)|^2)=1$, and
$|c_{\pm}(k)\rangle$, $\langle c_{\mp}(k)|c_{\pm}(k)\rangle=0$ are
the coin states corresponding to the eigenvalues $\lambda_{\pm}(k)$.
Since $\langle c_{-}(k)|\Delta X|c_{+}(k)\rangle=\langle
c_{+}(k)|\Delta X|c_{-}(k)\rangle^{\ast} = g(k)$, the expectation
value of $\Delta X(t)$ may be written as $A+B(t)$, where $A$ is a
constant corresponding to the uniform velocity of the wave packet,
and $B(t)$ is a time dependent term
\begin{equation} \label{e9}
B(t)=\int_{-\pi}^{+\pi} dk \left(
f_{+}^{\ast}(k)f_{-}(k)g(k)(\lambda_{+}^{\ast}(k)\lambda_{-}(k))^t +
c.c. \right).
\end{equation}
The above vanish if the initial wave packet consists only of the
positive or the negative energy eigenfunctions. As before, one may
write the eigenvalues in the exponential form
$\lambda_{\pm}(k)=e^{-i\varphi_{\pm}(k)}=e^{-iE_{\pm}(k)\Delta t}$,
so the time dependent part under integral, Eq.~(\ref{e9}), yields
\begin{equation} \label{e10}
(\lambda_{+}^{\ast}(k)\lambda_{-}(k))^t =e^{i(E_+(k)-E_-(k))t},
\end{equation}
which has the form of oscillations. If we take the initial wave
packet largely spread over the position space (for example broad
Gaussian packet) and multiply it by the factor $e^{ik_{0}x}$, the
corresponding momentum wave packet would be mainly localized around
$k_0$. In this case Eq.~(\ref{e9}) can be approximated by
\begin{equation} \label{e11}
B(t) \approx B \cos\left((E_+(k_0)-E_-(k_0))t\right),
\end{equation}
where $B$ is a time independent amplitude of oscillations. Of
course, time is discrete and $t=n\Delta t$.

Zitterbewegung is very well visible if one chooses the standing wave
packet, what in the case of DTQW does not always mean $k_0=0$. It is
visible if one calculates the group velocity
\begin{equation} \label{e12}
v_{\pm}(k) = \frac{\partial E_{\pm}}{\partial k}=\frac{i}{\Delta
t}\lambda_{\pm}^{\ast}(k)\frac{\partial \lambda_{\pm}(k)}{\partial
k}.
\end{equation}
Without loosing generality we assume $\Delta t=1$ and obtain
\begin{equation} \label{e13}
v_{\pm}(k)
=\pm\frac{n_z\sin{\frac{\theta}{2}}\cos{k}-\cos{\frac{\theta}{2}}\sin{k}}
{\sqrt{1-(\cos{\frac{\theta}{2}}\cos{k}+n_z
\sin{\frac{\theta}{2}}\sin{k})^2}}.
\end{equation}
Since the wave packet represents particle which does not move,
$(E_+(k_0)-E_-(k_0))\Delta t$ has the meaning of the mass. This
result is closely related to the one of the Dirac equation with the
oscillation frequency approximating $2m$. For the Hadamard walk,
$v_{\pm}(k_0)=0$ for $k_0=\pi/2$ and $\lambda_{\pm}(k_0)=\frac{1\pm
i}{\sqrt{2}}$, thus $(E_+(k_0)-E_-(k_0))\Delta t=\pi/2$.

The presence of Klein's paradox and Zitterbewegung in DTQW leads to
the bunch of questions. First of all, are the two effects truly
relativistic, since DTQW might be implemented with non-relativistic
quantum mechanics? The implementations of DTQW has been presented
for various physical systems, from optical latices to trapped ions
\cite{Dur, Joo, Hillery, Jeong, Eckert, Trav}, just to point some of
them. One may even think of the Stern-Gerlach experiment as of a
quantum quincunx with the magnetic field along $x$ and $z$ axes at
even and odd time steps respectively. The implementation schemes
might be different, but it is worth to notice that none of them has
anything to do with the Dirac equation. Of course one may wonder if
it is right to consider such concepts like spin and talk about
non-relativistic quantum physics, but the fact is that we may
include spin in the non-relativistic Schr\"odinger equation without
bothering about its origin. Moreover, this equation would correctly
predict the behavior of a real system.

In conclusion, effects associated with relativistic quantum
mechanics were shown to appear in non-relativistic models. We
derived formula for Zitterbewegung oscillations, Eq.~(\ref{e11}),
and presented in graphical way the nature of Klein's paradox. DTQW
was also compared with the discrete Dirac equation and it was shown
that the two models are related to each other and that one may go
from one to another by changing the reference frame.

The author would like to thank Micha{\l} Kurzy\'nski, Antoni
W\'ojcik and Andrzej Grudka for their kind help, stimulating
conversations and pointing important references.


\begin{thebibliography}{99}
\expandafter\ifx\csname natexlab\endcsname\relax\def\natexlab#1{#1}\fi
\expandafter\ifx\csname bibnamefont\endcsname\relax
  \def\bibnamefont#1{#1}\fi
\expandafter\ifx\csname bibfnamefont\endcsname\relax
  \def\bibfnamefont#1{#1}\fi
\expandafter\ifx\csname citenamefont\endcsname\relax
  \def\citenamefont#1{#1}\fi
\expandafter\ifx\csname url\endcsname\relax
  \def\url#1{\texttt{#1}}\fi
\expandafter\ifx\csname urlprefix\endcsname\relax\def\urlprefix{URL }\fi
\providecommand{\bibinfo}[2]{#2}
\providecommand{\eprint}[2][]{\url{#2}}

\bibitem[{\citenamefont{Ahar}(1993)}]{Ahar}
\bibinfo{author}{\bibfnamefont{Y.}~\bibnamefont{Aharonov}},
\bibinfo{author}{\bibfnamefont{L.}~\bibnamefont{Davidovich}},
and \bibinfo{author}{\bibfnamefont{N.}~\bibnamefont{Zagury}},
  \bibinfo{journal}{Phys. Rev. A},
  \bibinfo{volume}{{\bf{48}}}, \bibinfo{page}{1687} (\bibinfo{year}{1993}).

\bibitem[{\citenamefont{Farhi}(1998)}]{Farhi}
\bibinfo{author}{\bibfnamefont{E.}~\bibnamefont{Farhi}} and
\bibinfo{author}{\bibfnamefont{S.}~\bibnamefont{Gutmann}},
  \bibinfo{journal}{Phys. Rev. A},
  \bibinfo{volume}{{\bf{58}}}, \bibinfo{page}{915} (\bibinfo{year}{1998}).

\bibitem[{\citenamefont{Kempe}(2003)}]{Kempe}
\bibinfo{author}{\bibfnamefont{J.}~\bibnamefont{Kempe}},
  \bibinfo{journal}{Contemp. Phys.},
  \bibinfo{volume}{{\bf{44}}}, \bibinfo{page}{307} (\bibinfo{year}{2003}).

\bibitem[{\citenamefont{Ambainis}(2004)}]{Ambainis}
\bibinfo{author}{\bibfnamefont{A.}~\bibnamefont{Ambainis}},
  \bibinfo{journal}{quant-ph/0403120}.

\bibitem[{\citenamefont{Wojc}(2003)}]{Wojc}
\bibinfo{author}{\bibfnamefont{D.K.}~\bibnamefont{W\'ojcik}} and
\bibinfo{author}{\bibfnamefont{J.R.}~\bibnamefont{Dorfman}},
  \bibinfo{journal}{Phys. Rev. Lett.},
  \bibinfo{volume}{{\bf{90}}}, \bibinfo{page}{230260} (\bibinfo{year}{2003}).

\bibitem[{\citenamefont{My}(2003)}]{My}
\bibinfo{author}{\bibfnamefont{A.}~\bibnamefont{W\'ojcik}},
\bibinfo{author}{\bibfnamefont{T.}~\bibnamefont{{\L}uczak}},
\bibinfo{author}{\bibfnamefont{P.}~\bibnamefont{Kurzy\'nski}},
\bibinfo{author}{\bibfnamefont{A.}~\bibnamefont{Grudka}} and
\bibinfo{author}{\bibfnamefont{M.}~\bibnamefont{Bednarska}},
  \bibinfo{journal}{Phys. Rev. Lett.},
  \bibinfo{volume}{{\bf{93}}}, \bibinfo{page}{180601} (\bibinfo{year}{2004}).

\bibitem[{\citenamefont{Str1}(2006)}]{Str1}
\bibinfo{author}{\bibfnamefont{F.W.}~\bibnamefont{Strauch}},
  \bibinfo{journal}{Phys. Rev. A},
  \bibinfo{volume}{{\bf{73}}}, \bibinfo{page}{054302} (\bibinfo{year}{2006}).


\bibitem[{\citenamefont{Bracken}(2006)}]{Bracken}
\bibinfo{author}{\bibfnamefont{A.J.}~\bibnamefont{Bracken}},
\bibinfo{author}{\bibfnamefont{D.}~\bibnamefont{Ellinas}},
and \bibinfo{author}{\bibfnamefont{I.}~\bibnamefont{Smyrnakis}},
  \bibinfo{journal}{quant-ph/06105195}.

\bibitem[{\citenamefont{Fynmann}(1965)}]{Feynmann}
\bibinfo{author}{\bibfnamefont{R.P.}~\bibnamefont{Feynmann}} and
\bibinfo{author}{\bibfnamefont{A.R.}~\bibnamefont{Hibbs}},
  \bibinfo{book}{{\it{Quantum Mechanics and Path Integrals}}},
  \bibinfo{}{McGraw-Hill, New York} (\bibinfo{year}{1965}).

\bibitem[{\citenamefont{Asl}(2004)}]{Asl}
\bibinfo{author}{\bibfnamefont{C.}~\bibnamefont{Aslangul}},
  \bibinfo{journal}{quant-ph/0406057}.

\bibitem[{\citenamefont{Wess}(1999)}]{Wess}
\bibinfo{author}{\bibfnamefont{P.P.F.}~\bibnamefont{Wessels}},
\bibinfo{author}{\bibfnamefont{W.J.}~\bibnamefont{Caspers}},
and \bibinfo{author}{\bibfnamefont{F.W.}~\bibnamefont{Wiegel}},
  \bibinfo{journal}{Europhys. Lett.},
  \bibinfo{volume}{{\bf{46}}(2)}, \bibinfo{page}{123-126} (\bibinfo{year}{1999}).

\bibitem[{\citenamefont{Sakurai}(1967)}]{Sakurai}
\bibinfo{author}{\bibfnamefont{J.J.}~\bibnamefont{Sakurai}},
  \bibinfo{book}{{\it{Advanced Quantum Mechanics}}},
  \bibinfo{}{Addison-Wesley} (\bibinfo{year}{1967}).

\bibitem[{\citenamefont{Nayak}(2000)}]{Nayak}
\bibinfo{author}{\bibfnamefont{A.}~\bibnamefont{Nayak}} and
\bibinfo{author}{\bibfnamefont{A.}~\bibnamefont{Vishwanath}},
  \bibinfo{journal}{quant-ph/0010117}.

\bibitem[{\citenamefont{Dur}(2002)}]{Dur}
\bibinfo{author}{\bibfnamefont{W.}~\bibnamefont{D\"ur}},
\bibinfo{author}{\bibfnamefont{R.}~\bibnamefont{Raussendorf}},
\bibinfo{author}{\bibfnamefont{V.M.}~\bibnamefont{Kendon}} and
\bibinfo{author}{\bibfnamefont{H.-J.}~\bibnamefont{Briegel}},
  \bibinfo{journal}{Phys. Rev. A},
  \bibinfo{volume}{{\bf{66}}}, \bibinfo{page}{052319} (\bibinfo{year}{2002}).

\bibitem[{\citenamefont{Joo}(2006)}]{Joo}
\bibinfo{author}{\bibfnamefont{J.}~\bibnamefont{Joo}},
\bibinfo{author}{\bibfnamefont{P.L.}~\bibnamefont{Knight}} and
\bibinfo{author}{\bibfnamefont{J.K.}~\bibnamefont{Pachos}},
  \bibinfo{journal}{quant-ph/0606087}.


\bibitem[{\citenamefont{Hillery}(2003)}]{Hillery}
\bibinfo{author}{\bibfnamefont{M.}~\bibnamefont{Hillery}},
\bibinfo{author}{\bibfnamefont{J.}~\bibnamefont{Bergou}} and
\bibinfo{author}{\bibfnamefont{E.}~\bibnamefont{Feldman}},
  \bibinfo{journal}{quant-ph/0302161}.

\bibitem[{\citenamefont{Jeong}(2004)}]{Jeong}
\bibinfo{author}{\bibfnamefont{H.}~\bibnamefont{Jeong}},
\bibinfo{author}{\bibfnamefont{M.}~\bibnamefont{Paternostro}} and
\bibinfo{author}{\bibfnamefont{M.S.}~\bibnamefont{Kim}},
  \bibinfo{journal}{Phys. Rev. A},
  \bibinfo{volume}{{\bf{69}}}, \bibinfo{page}{012310} (\bibinfo{year}{2004}).

\bibitem[{\citenamefont{Eckert}(2005)}]{Eckert}
\bibinfo{author}{\bibfnamefont{K.}~\bibnamefont{Eckert}},
\bibinfo{author}{\bibfnamefont{J.}~\bibnamefont{Mompart}},
\bibinfo{author}{\bibfnamefont{G.}~\bibnamefont{Birkl}} and
\bibinfo{author}{\bibfnamefont{M}~\bibnamefont{Lewenstein}},
  \bibinfo{journal}{Phys. Rev. A},
  \bibinfo{volume}{{\bf{72}}}, \bibinfo{page}{012327} (\bibinfo{year}{2005}).


\bibitem[{\citenamefont{Trav}(2002)}]{Trav}
\bibinfo{author}{\bibfnamefont{B.C.}~\bibnamefont{Travaglione}} and
\bibinfo{author}{\bibfnamefont{G.J.}~\bibnamefont{Milburn}},
  \bibinfo{journal}{Phys. Rev. A},
  \bibinfo{volume}{{\bf{65}}}, \bibinfo{page}{032310} (\bibinfo{year}{2002}).


\end{thebibliography}
\end{document}